\documentclass[11pt]{article}
\usepackage{amsmath,amsfonts,epsfig,mathrsfs,todonotes,yfonts}
\usepackage{color}
\usepackage{cite}
\usepackage{stmaryrd}
\usepackage{scalerel}
\usepackage[top=2.5cm, bottom=2.5cm, left=2.75cm, right=2.75cm]{geometry} 
\usepackage{dsfont}
\numberwithin{equation}{section}

\newcommand{\+}{{+\!\!\!+}}
\newcommand{\one}{{\mathds{1}}}
\newcommand{\pa}{\partial} 
\newcommand{\ta}{{\tilde A}}
\newcommand{\taB}{{\tilde\bA}}
\newcommand{\bA}{\mathbf{A}}
\newcommand{\bAB}{{\bar\bA}}
\newcommand{\tb}{{\tilde B}}
\newcommand{\bB}{\mathbf{B}}
\newcommand{\bBB}{{\bar\bB}}
\newcommand{\tc}{{\tilde C}}
\newcommand{\td}{{\tilde D}}
\newcommand{\bI}{\mathbf{i}}
\newcommand{\bJ}{\mathbf{j}}
\newcommand{\bIB}{{\bar\bI}}
\newcommand{\bJB}{{\bar\bJ}}
\newcommand{\hal}{{\hat\al}}
\newcommand{\hatq}{{\hat q}}
\newcommand{\aB}{{\bar a}}
\newcommand{\bbR}{{\mathbb{R}}}

\newcommand{\then}{\Rightarrow}
\newcommand{\lrow}{\leftrightarrow}

\newcommand{\al}{\alpha}
\newcommand{\be}{\beta}
\newcommand{\ga}{\gamma}
\newcommand{\Ga}{\Gamma}
\newcommand{\de}{\delta}
\newcommand{\ka}{\kappa}
\newcommand{\vp}{\varphi}
\newcommand{\La}{\Lambda}
\newcommand{\la}{\lambda}

\newcommand{\tE}{\tilde{E}}
\newcommand{\tG}{\tilde{G}}

\newcommand{\ber}{\begin{eqnarray}}
\newcommand{\eer}[1]{\label{#1}\end{eqnarray}}
\newcommand{\eero}{\end{eqnarray}}

\newcommand{\nn}{\nonumber}
\newcommand{\na}{\nabla}
\newcommand{\nbar}{\bar\na}
\newcommand{\pbar}{\bar \pa}

\newcommand{\half}{{\textstyle{\frac12}}}
\newcommand{\ihalf}{{\textstyle{\frac i 2}}}

\newcommand{\Ka}{{K\"ahler}}
\newcommand{\bbX}[1]{\mathbb{X}^{#1}}
\newcommand{\smp}[1]{{\scaleto{({#1})}{7pt}}}
\newcommand{\bbXB}[1]{\bar{\mathbb{X}}^{#1}}
\newcommand{\bbD}[1]{\mathbb{D}_{#1}}
\newcommand{\bbDB}[1]{\bar{\mathbb{D}}_{#1}}
\newcommand{\re}[1] {(\ref{#1})}

\usepackage{lipsum}     

\begin{document}
\renewcommand{\theequation}{\thesection.\arabic{equation}}
\setcounter{page}{0}
\thispagestyle{empty}

\begin{flushright} \small
UUITP-08/20\\   Imperial-TP-2020-UL-02\\
YITP-SB-2020-7\
\end{flushright}
\begin{center} \LARGE
{\bf $\be\ga$-systems interacting with sigma-models}
\\[12mm] 
\normalsize
{\bf Ulf~Lindstr\"om$^{ab}$ and Martin Ro\v cek$^c$}\\
~\\
{\small\it
$^a$ The Blackett Laboratory, Imperial College London\\
Prince Consort Road, London SW7 2AZ, U.K.\\
~\\
$^b$Department of Physics and Astronomy
Uppsala University, \\ Box 516,
SE-751 20 Uppsala, Sweden \\
~\\
$^c$C.N.Yang Institute for Theoretical Physics, Stony Brook University, \\
Stony Brook, NY 11794-3840,USA\\}
~\\[1cm]
{\bf Abstract:}
\end{center}

We find a geometric description of interacting $\be\ga$-systems as a null Kac-Moody quotient of a nonlinear sigma-model for systems with varying amounts of supersymmetry.
\vskip8cm
{\footnotesize \noindent{\em emails:}
\newline
ulf.lindstrom@physics.uu.se
\newline
martin.rocek@stonybrook.edu}
\eject
\tableofcontents
\section{Introduction}

Generalized $\be\ga$-systems arise in many contexts -- including string theory and conformal field-theory; many papers have explored their quantum properties -- see, e.g.\cite{background}. In this paper, we explore the geometry of such systems interacting with general nonlinear sigma-models. We restrict our attention to left-moving $\be\ga$-systems, but the extension to include right-moving systems is straightforward. Our paper is only indirectly related to the work on chiral bosons -- see, e.g.\cite{chiralbosons}. 
After completing this work, the relevance of \cite{Witten:2005px} was pointed out to us -- it studies quantum and mathematical aspects of certain models related to the ones we describe here; our work focuses on a covariant geometric, albeit classical, description using (supersymmetric) sigma-models.

Consider a free $\be\ga$-system, that is a system with bosonic fields with a chiral 
action\footnote{Throughout this paper, we use $b,c$ for left-moving fields with integer spin (regardless of statistics), and $\be,\ga$ for their superpartners.}
\ber
S_b=\int d^2x\, b\pbar c~,
\eer{freebee}
which has field equations 
\ber
\pbar c=0~,~~~\pbar b=0~.
\eer{freom}
We assume that $b\equiv b_\+$ has spin one, and $c$ is a scalar. Clearly this system in not a sigma-model, and the target space is not a manifold in the usual sense. We can find a geometric description of this system as follows: we reinterpret $b$ as the gauge connection of a Kac-Moody symmetry on a certain manifold with indefinite signature.
We start with
\ber
\hat S=\int d^2x\, \pa \hat q \pbar c~,
\eer{freegeo}
which is a sigma-model with target space $\bbR^{1,1}$. This has a (right-moving) Kac-Moody symmetry\footnote{We thank Samson Shatashvili for pointing out that on curved world sheets, linear dilaton terms could lead to subtleties with this symmetry.}
\ber
\de \hat q =\la~,~~~\pa\la=0~
\eer{freek}
(Clearly, it also has a left-moving Kac-Moody symmetry, but we are not interested in it).
If we gauge this Kac-Moody symmetry by introducing a connection $b$
\ber
\pa \hat q \to \na\hat q:=\pa\hat q +b~,
\eer{freecoup}
we can choose a gauge $\hat q=0$, and the gauged version of \re{freegeo} reduces to 
\re{freebee}. We thus have found a geometric interpretation of our $\be\ga$-system: it is a chiral or Kac-Moody quotient along a null killing vector of a sigma-model with target space $\bbR^{1,1}$.

In this paper, we generalize this to interacting systems with various amounts of supersymmetry. Throughout this paper, we have assumed that the fields $b,c$ are commuting,  as $c$ corresponds to a coordinate on a target space manifold. However, very little changes if we let $b,c$ be anticommuting -- we are just studying sigma-models into a target supermanifold.

In Sec.\,2, we consider a broad class of generalized bosonic $\be\ga$-systems and find their geometric interpretation. In Sec.\,3, we repeat the exercise in $(1,1)$ superspace; the couplings to the fermions clearly reflect the underlying geometry in a nontrivial way. In Sec.\,4, we increase the supersymmetry to $(1,2)$; in this case the geometric sigma-model is a pseudo SKT geometry (strong \Ka~with torsion), and the chiral quotient is different from the usual $(1,2)$ quotient. In Sec.\,5, we describe the same system in $(2,1)$ superspace; in this case, the usual quotient gives the $\be\ga$-system. One significant difference is that left-moving $\be\ga$-systems are necessarily complex in $(1,2)$ superspace but {\em not} in $(2,1)$ superspace. In Sec.\,6, we consider $(2,2)$ superspace. In this case, these models arise naturally in terms of semichiral superfields, and we find a pseudo generalized \Ka~geometry. Finally, in Sec.\,7, we discuss our results and further possible developments. 

\section{Bosonic models}\label{babble}
In this section, we introduce the general bosonic sigma-model interacting with a commuting spin one left-moving $\be\ga$-system, and discuss its properties. We then find a geometric sigma-model whose quotient by a null symmetry gives the interacting $\be\ga$-system, and discuss its properties. Finally, we discuss various special cases of interest.

\subsection{Definitions and properties}
Let $E_{AB}=\half(G_{AB}+B_{AB})$ be the sum of the metric and the $B$ field, and consider
\ber
S=\int dx \Big(\pa\phi^AE_{AB}\pbar\phi^B+b_\al A^\al_B\pbar\phi^B\Big)~,
\eer{implag0}
where we combine the sigma-model fields $\phi^i$ with $c^\al$ and write a generic coordinate
\ber\{\phi^A\}\equiv\{\phi^i,c^\al\}~.
\eero
As long as it is invertible, we  can always choose  $A^\al_\be=\de^\al_\be$ by redefining $b$, which gives:
\ber
A^\al_B=\de^\al_B+\de^j_BA^\al_j(\phi)~~~\iff~~A^\al_B=(A^\al_j , \de ^\al_\be)~.~~~~ 
\eer{A1}
Then we can absorb $E_{B\al}$ by a shift of $b_\al$:
\ber
b_\al= b_\al'-\pa\phi^BE_{B\al}~,~\hbox{which leads to}~E_{Aj} = E_{Aj}'+E_{A\be}A^\be_j~.
\eer{bshift} 
Dropping the ${}'$, we are left with
\ber
E_{AB}=\left(E_{Ai}, 0\right)\equiv \left(   \begin{matrix} 
E_{ij} & 0 \\
E_{\al j} & 0 \\
\end{matrix}\right)~,
\eer{E1}
which we call the minimal frame.
The action \re{implag0} then reads
\ber
S=\int dx \Big(\pa\phi^AE_{Aj}\pbar\phi^j+b_\al \pbar c^\al+ b_\al A^\al_j \pbar\phi^j\Big)
\eer{implag}
The field equations that follow from extremizing \re{implag} are\footnote{Antisymmetrization is $A_{[i}B_{j]}:=A_iB_j-B_jA_i$ etc.}
\ber\label{beq}
&&\pbar c^\al+A^\al_i\pbar\phi^i=0~,\\[2mm]\nn
&& E_{\al j}\pa\pbar\phi^j+\pbar\phi^j\Ga^{(+)}_{jA\al}\pa\phi^A+\pbar b_\al-b_\be A^\be_j,_\al\pbar\phi^j=0~,\\[1mm]
&&G_{ij}\pa\pbar\phi^j+E_{\al i}\pa\pbar c^\al+\pbar\phi^B\Ga^{(+)}_{BAi}\pa\phi^A
-b_\al A^\al_{[j},_{i]}\pbar\phi^j+\pbar b_\al A^\al_i=0~,
\eer{feq1}
where we have used
\ber\nn
\left(E_{BA},_D+E_{AD},_B-E_{BD},_A\right)\!\!&\!\!=\!\!&\!\!\half\left(G_{BA},_D+G_{AD},_B-G_{BD},_A+B_{BA},_D+B_{AD},_B-B_{BD},_A\right)\\
\!\!&=:\!\!&\!\Ga^{(0)}_{BDA}-\half H_{BDA}=:\Ga^{(-)}_{BDA}=\Ga^{(+)}_{DBA}~,
\eer{gammas}
Part of our purpose is to find a geometric interpretation of these equations, which we do below.

We now discuss the formal symmetries of the action \re{implag}. We expect these to include diffeomorphisms and $B$ field gauge transformations, modified so that they preserve the minimal form of $E$ in \re{E1}. To this end we note that the action \re{implag} is invariant under two symmetries which do not preserve \re{E1}, and therefore can be used as compensating transformations to restore the minimal frame. The first does not transform the coordinates:
\ber
\de\phi^A=0~,~~~\de E_{AB}=-\ka_{A\al}A^\al_B~,~~~\de b_\al=\ka_{A\al}\pa\phi^A~.
\eer{X}
The second is any transformation that preserves the sigma-model term in the action and transforms the rest as
\ber
\de\left(A^\al_A\pbar\phi^A\right)=-\mu^\al_\be\left(A^\be_A\pbar\phi^A\right)~,~~~\de b_\al=b_\be \mu^\be_\al~.
\eer{Y}

The  B-field transformation
\ber
\de_{\bf B} E_{AB}=\half\de_{\bf B} B_{AB}\equiv\pa_A\La_B-\pa_B\La_A
\eero
preserves the action but not the form of $E$ \re{E1}. To restore the form we add a $\ka$-transformation \re{X} with parameter 
\ber
\ka_{A\al}= \pa_{[A}\La_{\al ]}
\eero
which implies
\ber
\de E_{A\al}=\pa_{[A}\La_{\al ]}
-\ka_{A\be}A^\be_\al=\pa_{[A}\La_{\al ]}
-\ka_{A\al}=0~,
\eero
as required.

Thus we find
\ber
&&\de_{\bf B} b_\al = \pa\La_\al-\pa\phi^A\pa_\al\La_A \nn\\[1mm]\nn
&&\de_{\bf B} E_{Aj}=(\pa_A\La_B-\pa_B\La_A)P^B_j\\[1mm]
&&\de_{\bf B} A^\al_i=0
\eer{modbees}
where the operator
\ber
P^A_j=\de^A_j-\de^A_\al A^\al_j
\eero
satisfies
\ber
(A^\al_j , \de ^\al_\be)\left(\begin{array}{c} \de^j_i\\ -A^\be_i\end{array}\right)\equiv A^\al_BP^B_i=0~,
\eero

The reparametrization symmetries\footnote{Note that the first term term in \re{re1} is cancelled by $ E_{AB},_C\de\phi^C$ for $\de\phi^C$ in \re{re0}.}
\ber\label{re0}
&&\de\phi^A=-\xi^A~,\\[1mm]
&&\de E_{Aj}=\xi^B\pa_BE_{Aj}+(\pa_A\xi^B)E_{Bj}+(\pa_j\xi^k)E_{Ak}
\eer{re1}
preserve the sigma-model part of the action \re{implag} but not the form of $E$ \re{E1}. To restore the form of $E$, we use a $\ka$-transformation \re{X}.  Since the second term in \re{implag} depends on $\phi^A$, we also need a  $\mu$ transformation \re{Y} to make the action invariant. The parameters are
\ber
\ka_{A\al}=\pa_\al \xi^j E_{Aj}~,~~~\mu^\be_\al=(\pa_\al\xi^B) A_B^\be~.
\eero
Since $E_{A\al}=0$, we need to check that its variation vanishes; using $E_{A\al}=0$, we find
\ber\nn
\de E_{A\al}&\!\!=&\!\!\xi^B\pa_BE_{A\al}+(\pa_A\xi^B)E_{B\al}+(\pa_\al\xi^B)E_{AB}-\ka_{A\be}A^\be_\al\\
&\!\!=&\!\!(\pa_\al\xi^B) E_{AB}-\ka_{A\al}=(\pa_\al\xi^\be) E_{A\be}=0~.
\eero
Thus we find
\ber\nn
&&\de b_\al=\pa\phi^B E_{Bj}\pa_\al \xi^j+ b_\be A^\be_B\pa_\al\xi^B\\[1mm]\nn
&&\de E_{Aj}=\xi^B\pa_BE_{Aj}+(\pa_A\xi^B)E_{Bj}+P^B_j(\pa_B\xi^k)E_{Ak}\\
&&\de A^\al_i=P^B_i\left(\pa_B\xi^\al+A^\al_j\pa_B\xi^j\right)+\xi^B\pa_BA^\al_i~.
\eer{moddiff}

\subsection{The Bosonic Geometric Model}
\label{linear}
To understand the geometry of the model, we use the same strategy as in \cite{Lindstrom:2007qf}: We think of  $b_\al$ as a connection and the term
\ber
b_\al A^\al_A\pbar\phi^A
\eero
as a gauge fixed version of
\ber
D\hatq_\al A^\al_A\pbar\phi^A=(\pa\hatq_\al+b_\al) A^\al_A\pbar\phi^A~.
\eer{bg}
This identifies $A^\al_A$ as the sum of metric and $B$-field
\ber
\pa \hatq_\al A^\al_A\pbar\phi^A=:\pa \hatq_\al \tE^\al_A\pbar\phi^A
\eero
in the ungauged sigma-model with additional coordinates $\hatq_\al$. The resulting geometry has a Kac-Moody isometry\footnote{The gauging of Kac-Moody isometries is discussed in\cite{Rocek:1991ps}.}: 
$\frac {\pa E}{\pa \hatq_\al}=0$.

The Lagrangian for this extended (ungauged) model is
\ber
\tilde L= \pa\phi^\ta\tE_{\ta \tb} \pbar\phi^\tb
\eero   
where 
\ber
\{\phi^\ta\}:= \{\phi^A,\hatq_\al\}=\{\phi^i,c^\al,\hatq_\al\}:=\{\phi^i,c^\al,\hatq^\hal\}~,
\eero
where we have introduced $\hatq^\hal:=\hatq_\al$ for convenience.
In general $E_{\ta \tb}$ is given by
\ber
\tE_{\ta \tb}\equiv \left(   \begin{matrix} 
E_{AB} & 0 \\
A^\al_B & 0
\end{matrix}\right)~,
\eer{egen}
which gives rise to the metric
\ber
\tG_{\ta \tb}\equiv \left(   \begin{matrix} 
G_{AB} & A^\be_A \\
A^\al_B & 0
\end{matrix}\right)~.
\eer{mgen}
The nonzero components of the connections $\tilde\Ga^{(+)}_{\ta\tb\tc}$ are
\ber\nn
&&\tilde\Ga^{(+)}_{AB\hat\ga}=A^\ga\!\!{}_A,_B\\[1mm]\nn
&&\tilde\Ga^{(+)}_{A\hat\be C}=A^\be\!{}_{[C},_{A]}\\[1mm]
&&\tilde\Ga^{(+)}_{ABC}=\Ga^{(+)}_{ABC}~.
\eer{ggen}

\subsection{The minimal frame}\label{miniable}
In the particular frame \re{A1},\re{E1} the matrix \re{egen} reduces to
\ber
\tE_{\ta \tb}\equiv \left(   \begin{matrix} 
E_{ij} & 0 & 0\\
E_{\al j} & 0 & 0 \\
A^\al_j & \de^\al_\be & 0
\end{matrix}\right)~,
\eero
and we note that $\tE_{\hal B}P^B_i=0$. The corresponding metric is 
\ber
\tilde G_{\ta \tb}= \left(   \begin{matrix} 
G_{ij} &  E_{\be i}  & A^\be_i\\
E_{\al j} & 0 & \de_\al^\be\\
A^\al_j & \de^\al_\be & 0
\end{matrix}\right)~, 
\eer{gtil}
which in general is invertible:
\ber
\tilde G^{\tb \ta}= \left(   \begin{matrix} 
\tG^{ji} &  -\tG^{kj}A^\al_k  & -\tG^{kj}E_{ \al k}\\
-A^\be_k\tG^{ki}& A^\be_k\tG^{kj} A^\al_j& \tG_\al^\be\\
-E_{ \be k}\tG^{ki}& \tG^\al_\be & E_{ \be k}\tG^{kj}E_{ \al j}
\end{matrix}\right)~.
\eero
Here 
\ber\nn
&&\tG^{ji}:=(G_{ij}-E_{ \al (i} A^\al_{j)})^{-1}\\[1mm]
&& \tG^\al_\be :=\de^\al_\be+\tG^{ij}E_{ \be i} A^\al_j~.
\eero
In particular, this implies that $\tilde G_{\ta\tb}$ is invertible in the general frame \re{egen}.  
We note that vectors of the form $(0,v^\al,0)$ and $(0,0,\hat v^\hal)$ are all null in the metric \re{gtil}. The metric (both in the minimal and the general frame) 
has signature $(n,k,-k)$ where 
$i=1\dots n,$ and $\al,\hal = 1\dots k$, as long as the interaction terms $E_{\al j},A^\al_j$ are not too large.

The field equations for the  extended sigma-model may be used to write those of the original model as follows
\ber\label{cbareq}
&&\pa (\tilde G_{\hal B}\pbar\phi^{B})=0\\[1mm]
&&\left[\tilde G_{ A\tb}\pa\pbar\phi^\tb+\tilde\Ga^{(+)}_{B\tc A}\pbar\phi^{B}\pa\phi^\tc\right]_{\pa\hatq^\hal=b_\al}=0
\eero
where \re{cbareq} is the derivative of \re{beq}, and we use
\ber\nn
&&\tilde\Ga^{(+)}_{jA\hal}=A^\al_j,_A\\[1mm]\nn
&&\tilde\Ga^{(+)}_{j\hal i}=A^\al_{[i},_{j]}\\[1mm]\nn
&&\tilde\Ga^{(+)}_{i\hal \be}=-A^\al_{i},_{\be}\\[1mm]
&&\tilde\Ga^{(+)}_{ABC}=\Ga^{(+)}_{ABC}~;
\eer{grel}
recall we use  $\hatq^\hal\equiv\hatq_\al$ for notational convenience. 

\subsection{Discussion}
We have seen that the model with the left-moving fields $b_\al,c^\al$ is a chiral quotient (Kac-Moody quotient) of a geometric sigma-model. We have assumed that 
$b_\al,c^\al$ are commuting, but aside from some obvious signs, the discussion would not change if some or all of them were anticommuting -- in that case the target space becomes a supermanifold, but the quotient proceeds in the same way.

In the general case \re{implag0}, for $E$ and $A$ to be functions of $c$, we require $c$ to be a scalar, and hence $b$ is a vector $b_\+$ on the world sheet. A particular special case arises when 
\ber
A^\al_B=A^\al,_B
\eer{Asym}
for some functions $A^\al$; then the second term in the action becomes
\ber
S_b=\int\, b_\al\pbar A^\al
\eer{semifree}
and the functions $A^\al$ are simply left-moving on-shell. We can change coordinates such that 
$c'{}^\al=A^\al(\phi,c)$. Then this term looks free, and all the interactions come through the dependence of $E$ on $c'$.

When \re{Asym} is satisfied, the connections \re{ggen} take a particularly simple form -- the nonvanishing components are: 
\ber\nn
&&\tilde\Ga^{(+)}_{AB\hat\ga}=A^\ga\!\!,_{AB}\\[1mm]
&&\tilde\Ga^{(+)}_{ABC}=\Ga^{(+)}_{ABC}~.
\eer{gsym}
When inserted into the definition of the curvature (\re{riem} below), the 
curvature has {\em no} components with hatted indices.

\section{$(1,1)$ Supersymmetry}\label{sable}
In this section we straightforwardly generalize the bosonic case -- both the interacting left-moving $\be\ga$-system
and the sigma-model whose quotient gives rise to it.
\subsection{The $(1,1)$ $\be\ga$-system}
The Lagrangian \re{implag0} is immediately generalized to $(1,1)$ superspace:
\ber
S=\int D_+D_-\Big[D_+\phi^AE_{AB}D_-\phi^B+\be_{\al +}A^\al_AD_-\phi^A\Big]~,
\eer{simplag}
where the scalars $\phi$ and the spinor $\be$ are $(1,1)$ superfields in representations of the supersymmetry algebra given in Appendix \ref{sp11}. As in the bosonic case, we combine the sigma-model fields $\phi^i$ with $c^\al$ and write a generic coordinate
\ber\{\phi^A\}\equiv\{\phi^i,c^\al\}~.
\eero
Again, we can chose the $E$ and $A$ in the special forms \re{E1} and \re{A1} using the same arguments to redefine $\be$.
Then the action has modified diffeomorphisms \re{moddiff} and $B$-field symmetries \re{modbees}. 

As above, when $A^\al_B=A^\al,_B$ is a gradient, the second term in the action simplifies to
\ber
S_\be=\int D_+D_-(\be_{\al +}D_-A^\al)
\eer{11semifree}
and the $\be$ field equation implies that the $A^\al$ are left-moving on shell:
\ber
D_-A^\al=0~~~\then
~~~D^2_-A^\al\equiv i\pbar A^\al=0~.
\eer{11semifreeeq}

To reduce \re{simplag} to components we shall need the following definitions\footnote{We now make the Lorentz vector structure of $b_\al$ manifest by writing $b_{\al\+}$~. Throughout, we define components of superfields by their spinor derivatives; it is not necessary to indicate a projection setting $\theta$'s to zero.}:
\ber\nn
\psi_\pm^A&:=&D_\pm\phi^A\\[1mm]\nn
F^A&:=&iD_+D_-\phi^A\\[1mm]\nn
\eta_{\al+}&:=&iD_+D_-\be_{\al +}\\[1mm]\nn
F_\al&:=&-iD_-\be_{\al +}\\[1mm]
b _{\al\+}&:=&-iD_+\be_{\al +}
\eer{betcomp}
The calculation of the component Lagrangian is straight forward albeit not very illuminating. In its place we follow the strategy of Sec.\,\ref{linear} to find the ungauged geometric Lagrangian and reduce that instead.

\subsection{The $(1,1)$ Geometric model}\label{unstable}

The Lagrangian for this higher-dimensional sigma-model is
\ber
\tilde L= D_+\phi^\ta\tE_{\ta \tb} D_-\phi^\tb
\eer{slag11}   
where the geometry is as in Sec.\,\ref{linear} with all fields now superfields. In particular, we have
\ber
\{\phi^\ta\}:= \{\phi^A,\hatq_\al\}=\{\phi^i,c^\al,\hatq_\al\}\equiv\{\phi^i,c^\al,\hatq^\hal\}~.
\eer{candq}
We define components as
\ber
\psi^\ta_+=D_+\phi^\ta,\quad 
\psi^\ta_-=D_- \phi^\ta,\quad F^\ta=iD_+D_-\phi^\ta\,.
\eer{ficomp} 
We collect terms and integrate by parts to get:
\ber\nn
&&S = \int d^2x \Bigg[\pa\phi^\ta E_{\ta\tb}\pbar\phi^\tb+\ihalf
(\psi_+^\ta\bar{\tilde\na}\psi_+^\tb+\psi_-^\ta\tilde\na\psi_-^\tb)\tG_{\ta\tb}\\\nn
&&\qquad\qquad -{\textstyle\frac 1 4} \tilde R^{(+)}_{\tc\td\ta\tb}\psi_+^\ta\psi_+^\tb\psi_-^\tc\psi_-^\td+\half \tG_{\ta\tb}
(F^\ta-i\Ga^{(+)\ta}_{\tc\td }\psi_+^\td\psi_-^\tc)(F^\tb-i\Ga^{(+)\tb}_{\tilde F\tE}\psi_+^{\tE}\psi_-^{\tilde F})
\Bigg]\\
\eer{act2}
where
\ber\nn
\bar{\tilde\na}\psi_+^\ta & =&\pbar\psi_+^\ta+\Ga^{(+)\ta}_{\tb \tc}\pbar\phi^\tb\psi_+^\tc\\[1mm]
\tilde\na\psi_-^\ta & =&
\pa\psi_-^\ta+\Ga^{(-)\ta}_{\tb \tc}\pa\phi^\tb\psi_-^\tc~.
\eer{cov}
Here $R^{(+)}_{\tc\td\ta\tb}$ is the Riemann curvature of $\Ga^{(+)}$:
\ber
\tilde R^{(+)}_{\ta\tb\tc\td}=\Ga^{(+)}_{[\tb|\tc\td|},^{}_{\ta]}
+\Ga^{(+)}_{[\ta |\tc\tE}\,\tilde G^{\tE\tilde F}\Ga^{(+)}_{|\tb] \td \tilde F}
\eer{riem}
Separating out the $i, \al$ and $\hat \al$ components is not particularly rewarding. However, we observe that it follows from the relations \re{ggen} and the fact that $\frac {\pa} {\pa \hatq^\hal}$ is an isometry, that the $\ta $ and $\tb$ indices of 
$R^{(+)}_{\ta\tb\tc\td}$ can never be $\hal$ or $\hat\be$. 

Since the metric $\tG_{\ta\tb}$ is invertible, we can eliminate the auxiliary fields $F^\ta$:
\ber
\tG_{\ta\tb}F^\ta=i\Ga^{(+)}_{\tc\td\tb}\psi^\td_+\psi^\tc_-~.
\eer{F0}
The details are given in the minimal frame in Appendix \ref{unable}.

The $\bar{\tilde\na}$-covariant derivatives in \re{cov} are
\ber
\tG_{\ta\tb}\bar{\tilde\na}\psi_+^\ta=\tG_{\ta\tb}\pbar\psi_+^\ta+\Ga^{(+)}_{\tc\ta\tb}\pbar\phi^\tc\psi^\ta _+
\eero
For $\tb=B$ this reads
\ber
\tG_{\ta B}\bar{\tilde\na}\psi_+^\ta=G_{AB}\nbar\psi^A_++\pbar(A^\al_B\psi^\hal_+)
-A^\al\!\!{}_C,_B\pbar\phi^C\psi^\hal_+~,
\eero
while $\tb=\hat\be$ yields
\ber
\tG_{\ta \hat\be}\bar{\tilde\na}\psi_+^\ta=\pbar(A^\be_A\psi^A_+)
+A^{\be}\!{}_{[C},_{A]}\pbar\phi^C\psi^A_+~.
\eero
Similarily we have for the ${\tilde\na}$ terms in  \re{cov}:
\ber
\tG_{\ta\tb}{\tilde\na}\psi_-^\ta=\tG_{\ta\tb}\pa\psi_-^\ta+\Ga^{(+)}_{\tc\ta\tb}\pa\phi^\ta \psi^\tc_-
\eero
For $\tb=B$ this reads
\ber
\tG_{\ta B}{\tilde\na}\psi_-^\ta=G_{AB}\na\psi^A_-
+A^\al_B\pa\psi^\hal_-+A^\al_{[B},^{}_{A]}\pa\phi^\hal\psi^A_-~,
\eero
and for $\tb=\hat\be$
\ber
\tG_{\ta \hat\be}{\tilde\na}\psi_-^\ta=\pa (A^\be_A\psi^A_-)~.
\eero

Using these formulae
we rewrite the action \re{act2} as
\ber\nn
&&S = \int d^2x \Bigg[\pa\phi^{ A}E_{ AB}\pbar\phi^{B}+\pa \phi^\hal A^\al_B\pbar\phi^B\\[-1mm]\nn
&&\qquad\qquad\qquad+i\Big\{
\half\psi_+^{A}G_{AB}\nbar\psi^B_+
+\psi_+^B\big[\pbar(A^\al_B\psi^\hal_+)
-A^\al\!\!{}_C,_B\pbar\phi^C\psi^\hal_+\big]\\[1mm]\nn
&&\qquad\qquad\qquad\qquad+\psi_+^{\hat\al}\big[\pbar(A^\al_B\psi^B_+)
+A^{\al}\!{}_{[C},_{B]}\pbar\phi^C\psi^B_+\big]+\half\psi_-^{A}G_{AB}\na\psi^B_-\\[1mm]\nn
&&\qquad\qquad\qquad\qquad\qquad\qquad\qquad\qquad
+\psi^B_-\big[A^\al_B\pa\psi^\hal_-+A^\al_{[B},^{}_{A]}\pa\phi^\hal\psi^A_-\big]\Big\}\\[-1mm]
&&\qquad\qquad\qquad -{\textstyle\frac 1 4} \tilde R^{(+)}_{CD\ta\tb}\psi_+^\ta\psi_+^\tb\psi_-^{C}\psi_-^{D}
\Bigg]~~.
\eer{gTasic}
To make contact with \re{simplag} we first gauge the Kac-Moody isometry $\frac {\pa} {\pa \hatq^\hal}$ by replacing (recall \re{candq} tells us $\phi^\hal\equiv\hatq^\hal$)
\ber
D_+\hatq^\hal \to \na_+\hatq^\hal:=D_+\hatq^\hal+\be_{\al +}~,
\eero
in analogy to \re{bg}, and choose a gauge where
\ber
\na_+\hatq^\hal\to\be_{\al +}~.
\eer{betaa}
Comparing the components of $\hatq^\hal$ from \re{ficomp} 
\ber
\phi^\hal=\hatq^\hal~,~~~\psi^\hal_+=D_+\hatq^\hal~,~~~\psi^\hal_-=D_-\hatq^\hal~,~~~F^\hal=iD_+D_-\hatq^\hal
\eero
to those of $\be_{\al +}$ in \re{betcomp} 
\ber
F_\al:=-iD_-\be_{\al +}~,~~~
b _{\al\+}:=-iD_+\be_{\al +}~,~~~
\eta^-_\al:=iD_+D_-\be_{\al +}
\eer{hoppsan}
we see from \re{betaa} that
\ber
\psi^\hal_+\to \be_{\al +}
\eero
in our gauge. With this identification it is clear that the auxiliary fields agree
\ber
F^\hal=F_\al~.
\eero
In addition we find from \re{hoppsan} that if we substitute $\be_{\al+}=D_+\hatq^\hal$, we get
\ber
b _{\al\+}=-iD_+D_+\hatq^\hal=\pa\hatq^\hal~,~~~\eta^-_\al=iD_+D_-D_+\hatq^\hal =\pa\psi^\hal_-~.
\eer{hoppat}
In the action, $\hatq^\hal$ and $\psi^\hal_-$ only appear in these combinations. We thus find the components of \re{simplag} with all $F$ auxiliary fields eliminated:
\ber\nn
&&S = \int d^2x \Bigg[\pa\phi^{ A}E_{ AB}\pbar\phi^{B}+b_{\al \+}A^\al_B\pbar\phi^B\\[-1mm]\nn
&&\qquad\qquad\qquad +
i\Big\{
\half\psi_+^{A}G_{AB}\nbar\psi^B_++\psi_+^B\big[\pbar(A^\al_B\be_{\al+})
-A^\al\!\!{}_C,_B\pbar\phi^C\be_{\al+}\big]\\[1mm]\nn
&&\qquad\qquad\qquad~~~+\be_{\al+}\big[\pbar(A^\al_B\psi^B_+)
+A^{\al}\!{}_{[C},_{B]}\pbar\phi^C\psi^B_+\big]
+\half\psi_-^{A}G_{AB}\na\psi^B_-\\[1mm]\nn
&&\qquad\qquad\qquad \qquad\qquad\qquad \qquad\qquad
+\psi^B_-\big[A^\al_B\eta_{+\al}+A^\al_{[B},_{A]}b_{\al \+}\psi^A_-\big]\Big\}\\[-1mm]\nn
&&\qquad\qquad -{\textstyle\frac 1 4}\Big( \tilde R^{(+)}_{CD A B}\psi_+^{A}\psi_+^{B}\psi_-^{C}\psi_-^{D}+2\tilde R^{(+)}_{CD A \hat\be}\psi_+^{A}\be_{\be +}\psi_-^{C}\psi_-^{D}+\tilde R^{(+)}_{CD \hal \hat \be}\be_{\al +}\be_{\be +}\psi_-^{C}\psi_-^{D}\Big)
\Bigg]\\
\eer{simplagg}
We note that $\eta$ is a fermionic auxiliary field whose equation is
$A^\al_B\psi^B_-=0$; this becomes $\psi^\al_-=-A^\al_j\psi^j_-$ in the minimal frame \re{E1},\re{A1}. Thus we have found a geometric form of the component action corresponding to \re{simplag}, including complicated interaction terms of the fermions.
We also observe that when $A^\al_B=A^\al\!\!,_B$ holds, the $b_\+,\be_+$ terms collapse to the component expansion of the semifree action \re{11semifree}:
\ber
S_\be=\int \Big[b_{\al \+}\pbar A^\al+i\be_{\al+}\pbar(A^\al_B\psi^B_+)\Big]~.
\eero

\section{ $(1,2)$ Supersymmetry}\label{mabelable}
For the bosonic and the $(1,1)$ models,  the relation between the sigma-model and its gauge-fixed reduction is straightforward. When we go to  $(1,2)$ supersymmetry, the natural extensions do not have the same clear relation. 
\subsection{The $(1,2)$ $\be\ga$-system}
Our starting point is the $(1,2)$ action for a $\be\ga$-system coupled to a sigma-model:
\ber\nn
&& S\,=i\int D_+\bbD{-}\bbDB{-}(k_AJ^A_BD_+\phi^B+\be_{\al +}J^\al_\be A^\be)\\[1mm]
&&~~\,= -\int D_+\bbD{-}\bbDB{-}(k_\bA D_+\phi^\bA-\bar k_\bAB D_+\bar\phi^\bAB+\be_{a +}A^a-\bar\be_{\aB+}A^{\aB})~,
\eer{s12red}
where we complexify all indices from the previous sections:
$\{A\}=\{\bA,\bAB\}$,  $\{\al\}=\{a,\aB\}$.
The $(1,2)$ superfields are $\{\phi^A\}\equiv\{\phi^\bA,\bar\phi^\bAB\}$, $\{\be_{\al+}\}\equiv\{\be_{a+},\bar\be_{\aB+}\}$, and obey the chirality conditions
\ber\nn
\bbDB{-}\phi^\bA=0&,&~~~\bbDB{-}\be_{a +}=0~~,\\
\bbD{-}\bar\phi^\bAB=0&,&~~~\bbD{-}\bar\be_{\aB+}=0~~.
\eer{chira}
The supersymmetry algebra is given in Appendix \ref{sp12}, and $J$ is a diagonal matrix such that $J^2=-\one$: it is $+i$ on holomorphic vectors and $-i$ on antiholomorphic vectors.

Reducing \re{s12red} to $(1,1)$ components, as described in Appendix \ref{sp12}, we find \re{simplag} with non-zero components:
\ber\nn
&&E_{\bA\bBB}= k^{}_{\bA},_{\bBB}~,~~~E_{\bAB\bB}=\bar k_{\bAB},^{}_\bB~,\\[1mm]
&&A^a_{\bBB}=A^a,_{\bBB}~,~~~A^{\aB
}_\bB=A^{\aB
},_\bB~,
\eer{eak}
where we have chosen a particular gauge for the $B$-field in $E$\cite{Hull:2008vw}.
More covariantly, we can write:
\ber
2E_{AB}=J_A^C\,k_C,_DJ^D_B+k_A,_B~~,~~~~2A^\al_B=J^\al_\ga \,A^\ga,_DJ^D_B+A^\al,_B~~.
\eer{cov12}

There are two ways we can satisfy $A^\al_B=A^\al\!\!,_B$ (cf.\,\re{Asym}) :
when $A^a,_\bB=0$, then $A^a$ is {\em anti}chiral: $\bbD-A^a=0$.  Then we can make a change of coordinates to replace $\bar c^{\aB}$ by $A^a$. The $\be$ equations of motion $\bbDB-A^a=0$ imply that $A^a$ is left-moving as in \re{11semifreeeq}; the complex conjugate works in the same way.

An alternative is to use a {\em real}\,\footnote{Clearly, we could choose $A^a$ equal to $A^{\bar a}$ up to a phase which can be absorbed by a redefinition of $\be$.} $A^\al$; since the $\be$ field equation implies $\bbDB-A^a=0$
and the $\bar\be$ field equation implies $\bbD-A^{\bar a}=0$, then $A^\al$ is left-moving.

In contrast to the previous cases in Secs.\,\ref{babble} and \ref{sable}, here we can only shift  $\be$ by chiral functions due to \re{chira}, which means we cannot choose the minimal form \re{E1} in $(1,2)$ superspace. 

\subsection{The $(1,2)$ Geometric model}\label{12gauging}
Alternatively, we start from a general $(1,2)$ sigma-model with isometries generated by 
$\frac\pa{\pa \hatq^{\hat a }}$:
\ber
S=-\int D_+\bbD{-}\bbDB{-}(k_\taB D_+\phi^\taB-\bar k_{\bar \taB }D_+\bar\phi^{\bar \taB })~,
\eer{s12geo}
where now 
\ber
\{\phi^\taB\}:= \{\phi^\bA,\hatq_a \}=\{\phi^\bI,c^a ,\hatq_a \}
~,~~~~~
\{\bar\phi^{\bar\taB}\}:= \{\bar\phi^{\bar \bA},\bar {\hatq}_{\aB
}\}
=\{\bar\phi^\bIB,\bar c^{\bar  a },\bar {\hatq}_{\aB
}\}
~.
\eero
Because of \re{eak}, the isometries
\ber
\frac\pa{\pa \hatq^{\hat a }}k_{\taB}=\frac\pa{\pa \hatq^{\hat a }}\bar k_{\bar\taB}= 0
\eero
(and their complex conjugates) imply that
$\tE$ has the form \re{egen} 
\ber
\tE_{\ta \tb}\equiv \left(   \begin{matrix} 
E_{AB} & 0 \\
A^\al_B & 0
\end{matrix}\right)~,
\eero

We could try to gauge the imaginary part of the isometries in chiral representation as described in \cite{Abou-Zeid:2019lgj}; in contrast to the
case of $(1,1)$ superspace above, this does not give the correct quotient model,  and so we need another procedure.

The key observation is that the action \re{s12geo} actually has a Kac-Moody symmetry: we can shift $\hatq_a\equiv\hatq^{\hat a }$ 
by right moving chiral parameters $\la^{\hat a }$ obeying
\ber
D_+\la=\pa\la=\bbDB- \la = 0
\eero
This can be promoted to a local symmetry with a $(1,2)$ chiral gauge parameter $\La^{\hat a }$ by introducing a novel {\em chiral} connection $\be^{\hat a }_+\equiv \be_{ a +}$ obeying $\bbDB-\be_{ a +}=0$, which gives
\ber
D_+\hatq_a  \to \na_+\hatq_a := D_+\hatq_a  +\be_{ a +}~,
\eero
where
\ber
\de \hatq_a  = \La_a ~,~~~\de\be_{ a +}=-D_+ \La_a ~,
\eero
and similarly for the complex conjugate. When we choose the gauge $\hatq=\bar{\hatq}=0$, we recover \re{s12red} with $A^a \equiv k_{\hat a }$. This is the correct complexified version of the $(1,1)$ story.

\section{ $(2,1)$ Supersymmetry}\label{mabelable2}
It is interesting to describe the same geometry in $(2,1)$ superspace. Here the description of the $\be\ga$-system is quite different; in particular,  as the complex structure appears in the opposite sector, there is no need to complexify the $\be\ga$-system. The quotient needed to descend from the geometric model to the $\be\ga$-system is the usual quotient \cite{Abou-Zeid:2019lgj}, as in the bosonic and $(1,1)$ cases.

\subsection{The $(2,1)$ $\be\ga$-system}
Our starting point is the $(2,1)$ action for a $\be\ga$-system coupled to a sigma-model; in this case, the form of the action appears geometric, but the ghost fields $c^\al$ are described by {\em unconstrained} scalar fields $X^\al$.
\ber\nn
&& S\,=i\int D_-\bbD+\bbDB+(k_iJ^i_jD_-\phi^j+k_\al D_-X^\al)\\[1mm]
&&~~\,= -\int D_-\bbD+\bbDB+(k_\bI D_-\phi^\bI-\bar k_\bIB D_-\bar\phi^\bIB-ik_\al D_-X^\al)~,
\eer{s21red}
where the indicies $\{i\}=\{\bI,\bIB\}$ are complexified.
The $(2,1)$ superfields are $\{\phi^i\}\equiv\{\phi^\bI,\bar\phi^\bIB\}$, and $\{X^\al\}$; 
the $\phi^i$ obey the chirality conditions
\ber
\bbDB+\phi^\bI=0~,~~~\bbD+\bar\phi^\bIB=0~,
\eer{chira21}
whereas $X^\al$ are unconstrained, and $J$ is a complex structure as in the previous section.
The supersymmetry algebra is given in Appendix \ref{sp21}.

Reducing \re{s21red} to $(1,1)$ components, as described in Appendix \ref{sp21}, we find \re{simplag} with non-zero components:
\ber\nn
&&2E_{ij}=J_j^n\,k_n,_mJ^m_i+k_j,_i~,~~2E_{\al i}=k_i,_\al~,~~2E_{i\al}=k_\al,_jJ^j_i~,
\\[1mm]
&&2A^\al_i\lrow
 k_j,_\al J^j_i-k_\al,_i~,~~2A^\al_\be\lrow
 k_\be,_\al-k_\al,_\be~,
\eer{cov21}
where the index mismatch for $A^\al_B$ arises because we identify 
$\be_\al\lrow\Psi^\al$; we also identify $X^\al\lrow c^\al$.

The condition $A^\al_B=A^\al\!\!,_B$ (cf.\,\re{Asym}) implies
\ber
k_j,_\al J^j_i=h_\al,_i~,~~~k_\be,_\al=h_\al,_\be~,
\eer{21asym}
where $h_\al$ is any real 1-form. Then
\ber
A^\al_B\lrow
 (h_\al-k_\al),_B~.
\eero
In $(2,1)$ superspace, this condition means that the equation of motion of $X^\al$ implies (cf.\,\re{11semifree})
\ber
D_- (h_\al-k_\al)=0~\then~\pbar  (h_\al-k_\al)=0~.
\eero

\subsection{The $(2,1)$ Geometric model}\label{21gauging}
In $(2,1)$ superspace, the geometric sigma-model is straightforward to find. 
Just as in \re{bg}, we identify $X$ as a connection gauging a symmetry of 
a general $(2,1)$ sigma-model by letting
\ber
X^\al \to X^\al + c^\al +\bar{ c}^\al
\eer{Xg}
where $ c$ is a chiral superfield:
\ber
\bbDB+ c=0 ~,~~~\bbD+\bar{ c}=0~.
\eero
Thus the ungauged geometric sigma-model is found by letting 
\ber
X^\al\to c^\al +\bar{ c}^\al
\eer{xqq}
and gives an action
\ber
S\,=i\int D_-\bbD+\bbDB+(k_\ta J^\ta_\tb D_-\phi^\tb)
\eer{s21geo}
where now 
\ber
\{\phi^\taB\}:= \{\phi^i, c^\al \}~,~~~~~
\{\bar\phi^{\bar\taB}\}:= \{\bar\phi^{\bar i},\bar { c}^\al\}~.
\eero

To compare to the $(1,1)$ geometric model, we need to interpret $c+\bar c$ as 
the real ghost field $c$ and $i(\bar c-c)$ as $\hatq$ in \re{candq}:
\ber
\{\phi^\ta\}:= \{\phi^A,\hatq_\al\}=\{\phi^i,c^\al,\hatq_\al\}~.
\eero
In this basis $\tE$ has the form \re{egen} 
\ber
\tE_{\ta \tb}\equiv \left(   \begin{matrix} 
E_{AB} & 0 \\
A^\al_B & 0
\end{matrix}\right)~,
\eero
with the components of $E$ and $A$ given \re{cov21}.

The sigma-model that we get after \re{xqq} has the obvious null isometry:
\ber
i\left(\frac\pa{\pa  c^\al}-\frac\pa{\pa \bar{ c}^\al}\right)~.
\eer{21iso}
This is actually a Kac-Moody symmetry, because $ c^\al +\bar{ c}^\al$ is invariant under
\ber
\de c^\al=i\la^\al~,~~~\bar{ c}^\al=-i\la^\al~~,~~~~\bbDB+\la=\bbD+\la=0\then
\pa\la=0~.
\eer{km21}
We can gauge the symmetry following \cite{Abou-Zeid:2019lgj} -- we start by introducing an unconstrained real scalar superfield $V$, which we identify with $X$ and let
\ber
 c^\al +\bar{ c}^\al\to X^\al + c^\al +\bar{ c}^\al~.
\eer{gauge21}
This combination is now gauge invariant under the complexified gauge transformations:
\ber 
\de c^\al=i\La^\al~,~~~\bar{ c}^\al=-i\bar\La^\al~,~~~\de X^\al=i(\bar\La^\al-\La^\al)~~,~~~~
\bbDB+\La=\bbD+\bar\La=0
\eer{21ginv}
Because only this combination enters in the gauged action, the gauge connection $\Ga_-$ does not appear in the action. Hence when we choose the gauge $ c=\bar{ c}=0$, we recover \re{s21red}.

\section{$(2,2)$ Supersymmetry}


We now consider $(2,2)$ superspace and find the relation to both $(1,2)$ superspace and $(2,1)$ superspace. To consider both left and right moving interacting $\be\ga$-systems, we need to consider such models.

\subsection{Models with only right semichirals}
\label{leftonly}
As pointed out in \cite{Buscher:1987uw} a model with only right semichiral fields describes a multiplet of free left moving bosons and left moving fermions.
Here we briefly recapitulate this. We use a notation consistent with the previous sections of this paper, albeit differing from the literature on semichiral multiplets 
\cite{Lindstrom:2005zr} and label the right semichiral fields by indices 
$\{\al\}\equiv\{a,\aB
\}$:
\ber
\bbDB-\bbX{a}=0~,~~~\bbD-\bbXB{\aB
}=0 
\eero
The $(2,2)$ action is 
\ber
S=\int \bbD{}^2\bbDB{}^2 K(\bbX{},\bbXB{})~.
\eer{Lact1}
The $(2,2)$ field equations that follow from this are
\ber
&&\bbDB{-} K_a=K_{a\bar b }~\bbDB{-}\bbXB{\bar b }=0~~~
\then
~~~ \bbDB{-} \bbXB{\aB
}=0~,
\eer{Fe1}
and the complex conjugate\footnote{Throughout this section, we use the abbreviation $K_a:=K,_a$, etc.}. In the last equality we assume that $K_{a\bar b }$ is invertible.
Using the results of Appendix \ref{spaceball}, we find that \re{Fe1} corresponds to the  $(1,1)$ equations:
\ber
D_{-}\bbX\al=0~,~~~ D_{-}\Psi^\al_+=0~~~\then
~~~
\pbar\bbX{\al }=\pbar\Psi^{\al}_+=0~,
\eer{Fe3}
where $\Psi^\al_+:=- J^\al_\be Q_+ \bbX\be$.

\subsection{Semichiral superfields interacting with sigma-models}\label{ablemable}
We now consider the action
\ber
S=\int \bbD{}^2\bbDB{}^2 K(\vp^i, \bbX{\al})~,
\eer{Lact2}
where $\vp^i$ are $(2,2)$ chiral $\Phi$ and/or twisted chiral $\chi$ superfields\footnote{This is a not the most general $(2,2)$ sigma-model -- for that, we would need to include further semichiral superfields of both chiralities. In this paper, we restrict our attention to the simpler case.}.

\subsubsection{Reduction to $(1,2)$ superspace}\label{stable}
To understand the geometry, we reduce to $(1,2)$ superspace and use the results of the Sec.\,\ref{mabelable}. The $(2,2)$ superfields 
$\vp,\bar\vp$ 
\ber
\{\vp^\bI\}=\{\Phi,\chi\}~,~~~\{\bar\vp^\bIB\}=\{\bar\Phi,\bar\chi\}
\eer{jmi} 
are holomorphic (resp.\ antiholomorphic) with respect to the complex structure $J_{\smp+}$:
\ber
J^{\,\bI}_{\!\smp+\bJ}\,d\vp^\bJ=i\,d\vp^\bI~,~~J^{\,\bIB}_{\!\smp+\bJB}\,d\bar\vp^\bJB=-i\,d\bar\vp^\bIB~.
\eer{jp}
Along with the right-chiral superfields $\bbX{},\bbXB{}$ these are identified with the $(1,2)$ superfields $\phi,\bar\phi$ as follows
\ber
\{\phi^\bA\}:=\{\phi^\bI,\phi^a\}=\{\Phi,\bar\chi,\bbX{}\}~,~~~\{\bar\phi^\bAB\}=\{\bar\phi^\bIB,\bar\phi^{\aB
}\}=\{\bar\Phi,\chi,\bbXB{}\}~,
\eer{jmh}
and are holomorphic (resp.\ antiholomorphic) with respect to the complex structure $J_{\smp-}$:
\ber
J^{\,\bA}_{\!\smp-\bB}\,d\phi^\bB=i\,d\phi^\bA~,~~J^{\,\bAB}_{\!\smp-\bBB}\,d\bar\phi^\bBB=-i\,d\bar\phi^\bAB~.
\eer{jm}
note that $\bar\chi$ is antichiral with respect to $J_{\smp+}$ and chiral with respect to  $J_{\smp-}$.
We emphasize that because the fields $\vp$ include chiral and twisted chiral fields but no semichiral fields, the complex structures $J_{\smp\pm}$ commute with each other \cite{Gates:1984nk}. When reduced to $(1,2)$ superpace \cite{Hull:2012dy}, as described in Appendix \ref{sp22}, the action becomes
\ber
S_{(1,2)}=i\int D_+\bbD{-}\bbDB{-}(K_i\,J^{\,i}_{\!\smp+j}\,D_+\vp^j+\Psi^\al_+K_\al)~,
\eer{s22red}
where $\Psi^\al_+:= Q_+ \bbX\be$ is $(1,2)$ chiral.
Comparing to \re{s12red}, we can identify
\ber
k_i=-K_j\,J^{\,j}_{\!\smp+k}\,J^{\,k}_{\!\smp-i}~,~~~ k_\al=0~,~~~\be_{\al+}\lrow
- J^\al_\be \Psi^\be_+~,~~~A^\al\lrow
 K_\al~, 
\eer{22to12}
where the different index positions on $\be_{\al+},A^\al$ relative to $\Psi^\al_+,K_\al$ arise because we use the usual convention for the coordinate $\bbX\al$.

Observe that when there is an isometry, e.g., when $K(\vp,\bar\vp,\bbX{}+\bbXB{})$, $A^a=A^{\bar a}$ as discussed below \re{chira}; then \re{Asym} is satisfied, and $A^\al$ is left-moving. This can be seen directly in $(2,2)$ superspace, as the $\bbX{},\bbXB{}$ field equations imply $\bbD-K_{\bbX{}}=\bbDB-K_{\bbX{}}=0$ (cf.~Sec.\,\ref{leftonly}).

We now substitute \re{22to12} into \re{cov12}; we must remember to identify $J_A^B$ from Sec.\,4 with $J_{\smp-}$. We then find the geometric quantities 
$E$ and $A$ which are used to write the $(1,1)$ superspace action:
\ber\nn
&&2E_{ij}=K^{}_{mn}\,J^{\,m}_{\!\smp+i}\,J^{\,n}_{\!\smp-j}-K_{mi}\,J^{\,m}_{\!\smp+n}\,J^{\,n}_{\!\smp-j}~,\\[1mm]\nn
&&2E_{\al i}=-K_{m\al}\,J^{\,m}_{\!\smp+n}\,J^{\,n}_{\!\smp-i}
~,~~2E_{i\al}=K_{j\be}\,J^{\,j}_{\!\smp+i}\,J^\be_\al~,
\\[1mm]
&&2A^\al_i\lrow K_{\al j}\,J^{\,j}_{\!\smp-i}-K_{\be i}\,J^\be_\al~,~~
2A^\al_\be\lrow K_{\al\ga}\,J^\ga_\be-K_{\be\ga}\,J^\ga_\al~.
\eer{eak22}

\subsubsection{Reduction to $(2,1)$ superspace}
The reduction of the model to $(2,1)$ superspace is simpler. We use \re{22to21mess} and \re{jx} to find
\ber
S_{(2,1)}=i\int D_-\bbD+\bbDB+\Big[K_iJ^{\,i}_{\!\smp-j}\, D_-\phi^i
+K_\al J^\al_\be D_- \bbX\be\Big]~.
\eer{22to21s}
Here $\phi^i$ are (anti)chiral $(2,1)$ superfields, $J_{\!\smp-}$ is as discussed in Sec.\,\ref{stable}, and $\bbX{}$ are complex unconstrained $(2,1)$ superfields. To compare to Sec.\,5,
we could decompose them into their real and imaginary parts, but it is more convenient to keep the complex coordinates.  We need to recall the $J^i_j$ in Sec.\,5 is now $J_{\smp+}$. Then we find
\ber
k_i=-K_j\,J^{\,j}_{\!\smp-k}\,J^{\,k}_{\!\smp+i}~,~~K_\al=K_\be\,J^\be_\al
\eer{22to21k}
Computing the $(1,1)$ quantities by substituting these into \re{cov21} gives exactly the same answer as above, namely \re{eak22}.

\subsection{The $(2,2)$ Geometric model}
To relate the $\be\ga$-system to a  $(2,2)$ sigma-model, we mimic the ALP  construction of \cite{Lindstrom:2007qf}. This is based on the interpretation of semichiral superfields as gauge fields for certain symmetries in a sigma-model with chiral and twisted chiral superfields.
We thus consider the action
\ber
S=\int \bbD{}^2\bbDB{}^2 K(\vp^i,X^\al)~,
\eer{Act2}
where
\ber
X^{a}:=\Phi^{a}+\bar\chi^{a} ~~,~~~\bar X^\aB:=\bar\Phi^\aB+\chi^\aB
\eer{Xdef}
with $\Phi$ and $\chi$ chiral and twisted chiral fields, respectively. The target space geometry is thus a torsionful geometry with a left and a right complex structure covariantly constant with respect to two torsionful connections\footnote {See, e.g., \cite{Lindstrom:2007qf}. In fact \re{Act2} is a special case of a chiral and twisted chiral sigma-model, and consequently, the left and right complex structures $J_{\smp\pm}$ commute.}.

The action is invariant under a complex Kac-Moody symmetry that preserves $X^{\al}$:
\ber
\de \Phi^a=\la^a~,~~~\de \bar\chi^a=-\la^a~,~~~
\de \bar\Phi^\aB=\bar\la^\aB~,~~~\de\chi^\aB=-\bar\la^\aB~,
\eer{KM1}
where
\ber\nn
\bbD+\la=\bbDB+\la=\pa\la=\bbDB-\la=0~,\\[1mm]
\bbD+\bar\la=\bbDB+\bar\la=\pa\bar\la=\bbD-\bar\la=0~.
\eer{KM2}
The quotient described below is analogous to what we found in Sec.\,\ref{12gauging}, namely a novel gauging for Kac-Moody symmetries.

To reduce to to $(1,2)$, we use
\ber\nn
&&Q_+X^{\al}=J^{\,\al}_{\!\smp+{\hat\be}}\,D_+Y^{\hat\be}~,\\[1mm]
&&Y^{\hat a}:=\Phi^a-\bar\chi^a~,~~~
\bar Y^{\bar{\hat a}}:=\bar\Phi^\aB-\chi^\aB~.
\eer{QXY}
We find \re{s12geo} with
\ber
k_\ta=-K_\tc\,\hat J^\tc{}_{\!\!\tb} \, J^{\,\tb}_{\!\smp-\ta}
\eer{22to12geo}
where $\hat J$ is $J_{\smp+}$ when written in a coordinates $\vp,X,Y$. Writing out the various indicies we have:
\ber
k_\bI=\pm K_\bI~,~~~k_{\hat a}=K_a~,~~~k_{a}=0~,
\eero
and similarly for the complex conjugates. The $\pm$ is $+$ for chiral superfields and $-$ for twisted {\em anti}chiral superfields, which are {\em both} chiral with respect to $J_{\smp-}$; see \re{jmh}.  Identifying $Y^\hal:=\hatq^\hal$, we recover
a special case of \re{s12geo}. 

Just as in the $(1,2)$ case, the standard gauging \cite{Lindstrom:2007vcc}
does not reduce the model 
to \re{Lact2}; instead, we gauge the Kac-Moody symmetry \re{KM1} as in 
\cite{Lindstrom:2007qf}. We introduce a right semichiral field $\bbX\al$
\ber
K(\vp^i,X^\al) ~\to~K(\vp^i,X^\al+\bbX{\al}) ~.
\eer{gag}
This potential is  now  invariant   under 
\ber
\de\phi^a=\La^a~,~~~\de\bar\chi^a=-\bar{\tilde\La}^a~,~~~
\de\bbX{a}=-\La^a+\bar{\tilde\La}^a
\eer{Lctf}
where $\La^a$ is chiral and $\bar{\tilde\La}^a$ is twisted antichiral.  Clearly we can then choose a gauge  where we gauge away $\phi^a,\bar\chi^a$; then
\ber
K(\vp^i,X^\al+\bbX{\al}) \to K(\vp^i,\bbX{\al}) 
\eer{gagu}
and we recover the form \re{Lact2}, now with knowledge about the underlying sigma-model geometry.  

\section{Discussion}
We have found a geometric way of understanding $\be\ga$-systems coupled to sigma-models with varying amounts of supersymmetry: as quotients along null Kac-Moody isometries of conventional sigma-models. 

We have studied the case with only left-moving $\be$ and $\ga$, and have only concerned ourselves with the classical geometric aspects -- in particular, we have not concerned ourselves with quantization and sigma-model anomalies, as discussed, e.g., in \cite{background},\cite{chiralbosons},\cite{Witten:2005px}. We expect the inclusion of right-moving $\be\ga$-systems to be straightforward; by describing left-moving $\be\ga$-systems in {\em both} $(1,2)$ and $(2,1)$ superspace, the methods to treat the right-moving systems are apparent.

For $(2,2)$ supersymmetric models, we have only considered sigma-models described by chiral and twisted chiral superfields; we expect the extension to the general case, including further left and right semichiral superfields, to be straightforward. Other superfield representations, namely complex linear and twisted complex linear superfields are equivalent to models with chiral and twisted chiral superfields.

It would be interesting to see if these considerations can be extended in any way to ``higher dimensional $\be\ga$-systems" \cite{Losev:1996up}.

\bigskip
\bigskip
\noindent{\bf\Large Acknowledgements}
\vskip1mm
\noindent We would like to thank S.James Gates, Chris Hull, Sergei Kuzenko, Samson Shatashvili, and Rikard von Unge for comments and suggestions. UL is grateful to the SCGP for hospitality and to Lars Hierta's foundation for partial support. MR thanks NSF-PHY-1915093 for partial support.

\appendix
\bigskip
\bigskip
\noindent{\bf\Large Appendices}
\vspace{-5mm}
\section{Superspaces}\label{spaceball}
In these appendices, we discuss the superspace for various superalgebras. Sigma-models 
have target space geometries that depend on the amount of supersymmetry.
For $(1,1)$, the geometry is (pseudo)Riemannian with a natural connection with torsion;
for $(1,2)$ or $(2,1)$, the geometry is (pseudo) strong \Ka~with torsion;
and for $(2,2)$, the geometry is (pseudo) generalized \Ka. 

\subsection{$(1,1)$ superspace}\label{sp11}
The $(1,1)$ superalgebra is generated by spinor derivatives $D_\pm$ that obey
\ber
D^2_+=i\pa~,~~~D^2_-=i\pbar~,~~~\{D_+,D_-\}=0~.
\eero
The $(1,1)$ superfields are unconstrained, and gauging is done with a spinor connection
$D_\pm\to\na_\pm=D_\pm+\be_\pm$.
The superspace action is written using the measure $D_+D_-$ as follows:
\ber
S:=\int d^2x\,D_+D_- ~L~.
\eer{11mess}

\subsection{$(1,2)$ superspace}\label{sp12}
The $(1,2)$ superalgebra is generated by the real spinor derivative $D_+$ and the complex spinor derivatives $\bbD-,\bbDB-$.
\ber
D_+^2=i\pa~,~~~ \{ \bbD{-}, \bbDB{-}\}=i\pbar~,~~~\{D_+,\bbD-\}=\{D_+,\bbDB-\}=0~.
\eer{12alg}
Right-(anti)chiral superfields obey $\bbDB-\phi=0, \bbD-\bar\phi=0$, resp.
Usual gauging involves a left-spinor connection $\be_+$ and a real potential $V$ -- see 
\cite{Abou-Zeid:2019lgj} for the details of the analogous $(2,1)$ case. 
As shown in Sec.\,\ref{12gauging}, we need a different kind of gauging that is suitable for Kac-Moody symmetries.

We reduce to $(1,1)$ using
\ber
\bbD{-}=\half(D_ -- iQ_-)~,~~~\bbDB{-}=\half(D_-+iQ_-)~,
\eer{12to11app}
from which it follows the superspace measure becomes
\ber
D_+\bbD-\bbDB-=\ihalf D_+D_-Q_-~~~.
\eer{12to11m}
When we push in $Q_-$ to find the $(1,1)$ action for chiral superfields, we use, e.g., 
\ber
Q_-\phi=iD_-\phi~,~~~Q_-\bar\phi=-iD_-\bar\phi~,
\eer{qphi}
which can be written covariantly for $\{\phi^i\}=\{\phi^\bI,\bar\phi^\bIB\}$ as
\ber
Q_-\phi^i=J^i_j\,D_-\phi^j
\eer{qphij}

\subsection{$(2,1)$ superspace}\label{sp21}
The $(2,1)$ superalgebra is generated by the real spinor derivative $D_-$ and the complex spinor derivatives $\bbD+,\bbDB+$.
\ber
D_-^2=i\pbar~,~~~ \{ \bbD{+}, \bbDB{+}\}=i\pa~,~~~\{D_-,\bbD+\}=\{D_-,\bbDB+\}=0~.
\eer{21alg}
Left-(anti)chiral superfields obey $\bbDB+\phi=0, \bbD+\bar\phi=0$, resp.
Usual gauging involves a left-spinor connection $\be_-$ and a real potential $V$ -- see 
\cite{Abou-Zeid:2019lgj} for the details. 
As shown in Sec.\,\ref{21gauging}, we need a different kind of gauging that is suitable for left Kac-Moody symmetries generated by parameters obeying $\pbar\la=0$.

We reduce to $(1,1)$ using
\ber
\bbD{+}=\half(D_ +- iQ_+)~,~~~\bbDB{+}=\half(D_++iQ_+)~,
\eer{21to11app}
from which it follows the superspace measure becomes
\ber
D_-\bbD+\bbDB+=-\ihalf D_+D_-Q_+~~~.
\eer{21to11m}
When we push in $Q_+$ to find the $(1,1)$ action for chiral superfields, we use, e.g., 
\ber
Q_+\phi=iD_+\phi~,~~~Q_+\bar\phi=-iD_+\bar\phi~,
\eer{qphip}
which can be written covariantly for $\{\phi^i\}=\{\phi^\bI,\bar\phi^\bIB\}$ as
\ber
Q_+\phi^i=J^i_j\,D_+\phi^j
\eer{qphijp}
On the other hand, for an unconstrained superfield $X$,  $Q_+X$ is independent as a $(1,1)$ superfield:
\ber
\Psi_+=Q_+X~.
\eer{21psi}

\subsection{$(2,2)$  superspace}\label{sp22}

The $(2,2)$ algebra of covariant derivatives is
\ber\nn
&\{\bbD+,\bbDB+\}= i\pa~,~~~~~~\{\bbD-,\bbDB-\}= i\bar\pa~,~~~~~~
\bbD{\pm}^2=0~,&\\[1mm]
&\{\bbD{+},\bbD{-}\}=0~,~~~~~~
\{\bbDB{\pm},\bbD{\mp}\}=0~,&
\eer{2alg}
and the complex conjugate relations.

Chiral superfields $\Phi^a$ satisfy:
\ber
\bar{\mathbb{D}}_{\pm}\Phi^a=\mathbb{D}_{\pm}\bar\Phi^{\aB
}=0~,
\eer{chir}
but in $d=2$ we may also introduce twisted chiral fields $\chi$ that satisfy
\ber
\bar{\mathbb{D}}_{+}\chi={\mathbb{D}}_{-}\chi=0~,~~
\mathbb{D}_{+}\bar\chi=\bar{\mathbb{D}}_{-}\bar\chi=0~.
\eer{Twist}
as well as left and right semichiral superfields; in this paper we only use\footnote{Usually, we write $\bbX{\ell},\bbXB{\bar\ell}$ for left semichiral fields (which obey $\bbDB+\bbX{\ell}=\bbD{+}\bbXB{\bar\ell}=0$) and $\bbX{r},\bbXB{\bar r}$ for right semichiral superfields \cite{Lindstrom:2005zr}; since here we use only right semichiral superfields, we drop their superscripts.}
right semichiral superfields which obey
\ber
\bbDB{-}\bbX{}=\bbD{-}\bbXB{}=0~.
\eer{LRdef}

To display the physical content we may rewrite an action in $(1,2)$ superspace.
By analogy to \re{12to11app}, we descend to $(1,2)$ superspace by defining the left-handed real spinor derivative
\ber
D_{+}\equiv \mathbb{D}_{+}+\bar{\mathbb{D}}_{+}~,
\eer{recov}
and the generator of second supersymmetry
\ber
Q_{+}\equiv i(\mathbb{D}_{+}-\bar{\mathbb{D}}_{+})~.
\eer{reduceop}
They satisfy
\ber
D^2_+=Q^2_+=i\pa~.
\eer{22to12algapp}
The $(2,2)$ measure reduces to
\ber
\bbD{}^2\bbDB{}^2:= -2\bbD+\bbD-\bbDB+\bbDB-=2\bbD+\bbDB+\bbD-\bbDB-
=iD_+\bbD-\bbDB-Q_+
\eer{22mess}
In  $(1,2)$ superspace, all superfields are either unconstrained or chiral; we now explain how $(2,2)$ superfields decompose into their $(1,2)$ components. From \re{chir}, we find
\ber
Q_+\Phi=JD_+\Phi
\eer{jP}
where $J$ is the canonical complex structure (diagonal $+i,-i$). Similarly, from \re{Twist}, we find
\ber
Q_+\chi=JD_+\chi~.
\eer{jchi}
However, $\Phi,\bar\chi$ are the $(1,2)$ {\em chiral} superfields, which we collectively denote as $\phi$. To distinguish $(2,2)$ and $(1,2)$ chirality properties, we use the notation $J_{\smp+}$ and $J_{\smp-}$ as explained in Sec.\,\ref{ablemable}. 

The right semichiral multiplets $\bbX{}$ give rise to two $(1,2)$ chiral multiplets: a scalar and a spinor:
\ber
\bbX~,~~\Psi_+=Q_+ \bbX~.
\eer{xpsi}

We can also reduce from $(2,2)$ to $(2,1)$ superspace. Everything proceeds analogously; in particular, we find
\ber
\bbD{}^2\bbDB{}^2=iD_-\bbD+\bbDB+Q_-~.
\eer{22to21mess}
The reduction of $(2,2)$ chiral and twisted chiral superfields to $(2,1)$ chiral superfields interchanges the roles of $J_{\smp+}$ and $J_{\smp-}$, but otherwise is unchanged; instead of \re{jP} and \re{jchi}, we find
\ber
Q_-\Phi=JD_-\Phi~~,~~~~Q_-\chi=-JD_-\chi~~.
\eero
However, in contrast to \re{xpsi}, right semichiral multiplets $\bbX{}$ now give rise to a complex {\em unconstrained} $(2,1)$ scalar superfield:
\ber
Q_-\bbX{}=JD_-\bbX{}~.
\eer{jx}

\section{Minimal frame components}\label{unable}
Here we work out the detailed form of various quantities in the minimal frame of Sec.\,\ref{miniable} (in particular, see \re{E1},\re{A1}). For the bosonic auxiliary field equations, when the indices $\tb=B$ in \re{F0}, the equations read
\ber
G_{AB}F^{A}+A^\al_B F_\al=i\Ga^{(+)}_{CDB}\psi^{D}_+\psi^{C}_ -- iA^\de_j,_B\psi^{\hat\de}_+\psi^{j}_-+iA^\de_k,_j\psi^{\hat\de}_+\psi^{j}_-\de^k_B~.
\eer{F1}
Choosing $B=\be$ and $B=j$ in turn in \re{F1} yields
\ber\nn
&&F^{\hat\be}=i\Ga^{(+)}_{CD\be}\psi^{D}_+\psi^{C}_ -- iA^\de_j,_\be\psi^{\hat\de}_+\psi^{j}_ -- E_{\be i}F^i\\[1mm]
&&G_{ij}F^i+A^\al_jF^\hal=i\Ga^{(+)}_{CDj}\psi^{D}_+\psi^{C}_ -- iA^\de_{[k},_{j]}\psi^{\hat\de}_+\psi^j_-~.
\eero
For $\tb=\hat\be$ \re{F0} reads
\ber\nn
&&\tG_{A\hat\be}F^A=A^\be_AF^A=i\tilde\Ga^{(+)}_{jA\hat\be}\psi^{A}_+\psi^{j}_-=iA^\be_j,_A\psi^{A}_+\psi^{j}_- \\[1mm]
&&\then
 ~F^\be=iA^\be_j,_A\psi^{A}_+\psi^{j}_- +A^\be_iF^i
\eero
The $\bar{\tilde\na}$-covariant derivatives in \re{cov} are
\ber
\tG_{\ta\tb}\bar{\tilde\na}\psi_+^\ta=\tG_{\ta\tb}\pbar\psi_+^\ta+\Ga^{(+)}_{\tc\td\tb}\pbar\phi^\tc\psi^\td_+
\eero
For $\tb=B$ this reads
\ber\nn
\tG_{\ta B}\bar{\tilde\na}\psi_+^\ta&\!\!=&\!\!G_{AB}\bar{\tilde\na}\psi_+^{A}+A^\al_B\bar{\na}\psi^\hal_+\\[1mm]
&\!\!=&\!\!G_{AB}\pbar\psi^A_+
+A^\al_B\pbar\psi^\hal_++\Ga^{(+)}_{CDB}\pbar\phi^C\psi^D_+-A^\de_i,_B\pbar\phi^i\psi^{\hat\de}_++A^\de_j,_i\pbar\phi^j\psi^{\hat\de}_+\de^i_B~,\nn\\
\eero
while $\tb=\hat\be$ yields
\ber
\tG_{\ta \hat\be}\bar{\tilde\na}\psi_+^\ta=A^{\be }_A\bar{\tilde\na}\psi^A_+=\pbar\psi^\be_++A^{\be}_i\pbar\psi^i_++A^{\be}_i,_D\pbar\phi^i\psi^D_+~.
\eero
Similarily we have for the ${\tilde\na}$ terms in  \re{cov}:
\ber
\tG_{\ta\tb}{\tilde\na}\psi_-^\ta=\tG_{\ta\tb}\pa\psi_-^\ta+\Ga^{(+)}_{\tc\td\tb}\bar\phi^\td\psi^\tc_-
\eero
For $\tb=B$ this reads
\ber\nn
&&\tG_{\ta B}{\tilde\na}\psi_-^\ta=G_{AB}{\tilde\na}\psi_-^{A}+A^\al_B{\na}\psi^\hal_-\\[1mm]
&&=G_{AB}\pa\psi^A_-
+A^\al_B\pa\psi^\hal_-+\Ga^{(+)}_{DCB}\pa\phi^C\psi^D_ -- A^\de_i,_B\pa\phi^{\hat\de}\psi^i_-+A^\de_j,_i\pa\phi^{\hat\de}\psi^j_-\de^i_B~,
\eero
and for $\tb=\hat\be$
\ber
\tG_{\ta \hat\be}{\tilde\na}\psi_-^\ta=A^\be_A{\tilde\na}\psi^A_-=\pa\psi^\be_-+A^\be_i\pa\psi^i_-+A^\be_i,_D\pa\phi^D\psi^i_-=\pa (A^\be_B\psi^B_-)~.
\eero

We next work out the details of the component action in the minimal frame.
\ber\nn
&&S = \int d^2x \Bigg[\pa\phi^{ A}E_{ Aj}\pbar\phi^{j}+\pa \phi^\hal A^\al_B\pbar\phi^B+
i\Big\{
\half\psi_+^{A}G_{AB}\nbar\psi^B_+\\[-1mm]\nn
&&\qquad\qquad\qquad +\psi_+^B[A^\al_B\pbar\psi^\hal_+
-A^\de_i,_B\pbar\phi^i\psi^{\hat\de}_+]
+\psi_+^iA^\de_j,_i\pbar\phi^j\psi^{\hat\de}_+\\[1mm]\nn
&&\qquad\qquad\qquad+\psi_+^{\hat\be}[\pbar\psi^\be_++A^{\be}_i\pbar\psi^i_++A^{\be}_i,_D\pbar\phi^i\psi^D_+]+\half\psi_-^{A}G_{AB}\na\psi^B_-\\[1mm]\nn
&&\qquad\qquad\qquad 
+\psi^B_-[A^\al_B\pa\psi^\hal_ -- A^\de_i,_B\pa\phi^{\hat\de}\psi^i_-]+\psi^i_-A^\de_j,_i\pa\phi^{\hat\de}\psi^j_-\Big\}\\[-1mm]
&&\qquad\qquad\qquad -{\textstyle\frac 1 4} \tilde R^{(+)}_{CD\ta\tb}\psi_+^\ta\psi_+^\tb\psi_-^{C}\psi_-^{D}
\Bigg]
\eer{Tasic}
To descend to the quotient model, we substitute 
\ber
\psi^\hal_+\to \be_{\al +}~,~~~b _{\al\+}:=-iD_+D_+\phi^\hal=\pa\phi^\hal~,~~~\eta^-_\al:=iD_+D_-D_+\phi^\hal =\pa\psi^\hal_-~.
\eer{apphop}
into \re{Tasic}; since 
$\phi^\hal$ and $\psi^\hal_-$ only appear as in \re{apphop}, this gives:
\ber\nn
&&S = \int d^2x \Bigg[\pa\phi^{ A}E_{ Aj}\pbar\phi^{j}+b_{\al \+}A^\al_B\pbar\phi^B+
i\Big\{
\half\psi_+^{A}G_{AB}\nbar\psi^B_+\\[-1mm]\nn
&&\qquad\qquad\qquad +\psi_+^B[A^\al_B\pbar\be_{\al+}
-A^\al_i,_B\pbar\phi^i\be_{\al+}]
+\psi_+^iA^\al_j,_i\pbar\phi^j\be_{\al+}\\[1mm]\nn
&&\qquad\qquad\qquad+\be_{\be+}[\pbar\psi^\be_++A^{\be}_i\pbar\psi^i_++A^{\be}_i,_D\pbar\phi^i\psi^D_+]+\half\psi_-^{A}G_{AB}\na\psi^B_-\\[1mm]\nn
&&\qquad\qquad\qquad 
+\psi^B_-[A^\al_B\eta_{+\al}-A^\al_i,_Bb_{\al\+}\psi^i_-]+\psi^i_-A^\al_j,_ib_{\al\+}\psi^j_-\Big\}\\[-1mm]\nn
&&\qquad\qquad -{\textstyle\frac 1 4}\Big( \tilde R^{(+)}_{CD A B}\psi_+^{A}\psi_+^{B}\psi_-^{C}\psi_-^{D}+2\tilde R^{(+)}_{CD A \hat\be}\psi_+^{A}\be_{\be +}\psi_-^{C}\psi_-^{D}+\tilde R^{(+)}_{CD \hal \bar \be}\be_{\al +}\be_{\be +}\psi_-^{C}\psi_-^{D}\Big)
\Bigg]\\
\eer{simplagc}
We note that $\eta$ is a fermionic auxiliary field whose equation
$A^\al_B\psi^B_-=0$ implies
\ber
\psi^\al_-=-A^\al_j\psi^j_-
\eero
since we are in the minimal frame.

\newcommand{\np}{{\em Nucl.\ Phys.\ }}
\newcommand{\pr}{{\em Phys.\ Rev.\ }}
\newcommand{\cmp}{{\em Commun.\ Math.\ Phys.\ }}
\newcommand{\pl}{{\em Phys.\ Lett.\ }}

\end{document}